\documentclass[preprint, epsf]{revtex4}
\usepackage{amssymb}

\renewcommand{\thefootnote}{\fnsymbol{footnote}}

\newcommand{\ba}{\begin{array}}
\newcommand{\ea}{\end{array}}
\newcommand{\bd}{\begin{displaymath}}
\newcommand{\ed}{\end{displaymath}}
\newcommand{\be}{\begin{equation}}
\newcommand{\ee}{\end{equation}}
\newcommand{\bea}{\begin{eqnarray}}
\newcommand{\eea}{\end{eqnarray}}

\newcommand{\pa}[1]{{\partial_{#1}}}

\def\k{\kappa}

\begin{document}
\preprint{BA-05-03}

\title{Hybrid Inflation, Dark Energy And Dark Matter}
\vskip 8cm
\author{Qaisar Shafi}
%\email{shafi@bartol.udel.edu}
\author{Arunansu Sil}
%\email{asil@bartol.udel.edu}
\author{Siew-Phang Ng}
%\email{spng@bartol.udel.edu}

\affiliation{Bartol Research Institute, University of Delaware,
Newark, DE 19716, USA}

\begin{abstract}
It has been suggested that the dark energy density $\rho_v \sim 
10^{-12}$ eV$^4$ in the
universe is associated with a metastable (false) vacuum, while the true
vacuum has a vanishing cosmological constant. By including supergravity
corrections we show how this is naturally realized in realistic supersymmetric
hybrid inflation models. With a fundamental supersymmetry breaking scale 
$\sim$ TeV,
the LSP is not a suitable candidate for cold dark matter. We consider axion
physics to overcome this and simultaneously provide a resolution of the 
MSSM $\mu$ problem.
\vskip 2cm 
\begin{center}
{\bf {In Memoriam Ib Arne Svendsen}}
\end{center}
\end{abstract}
\maketitle
%\newpage

\renewcommand{\thefootnote}{\arabic{footnote}}
\setcounter{footnote}{0}

Recent studies of the cosmic microwave background radiation \cite{cm},
Supernovae 1a \cite{snova} and large scale structure \cite{large}, 
taken collectively, 
present a fairly compelling case for a dark (vacuum) energy density 
$\rho_v \sim 10^{-12}$ eV$^4$.
Indeed, $\rho_v$ is estimated to provide almost 70$\%$ of the critical energy
density,
with matter (including baryons and possibly neutrinos) making up the remaining
30$\%$ or so. Understanding the origin of $\rho_v$ poses one of the most
fundamental
theoretical challenges, namely how $\rho_v \sim 10^{-120} M_P^4$ happens to 
be so much smaller than $M_P^4$, where $M_P =2.4 \times 10^{18}$ GeV 
denotes the reduced Planck mass. 
Another related problem is to understand how 
$\rho_v$ and the matter density $\rho_m$ which, in principle, can be 
expected to scale very differently with the universe expansion, are of 
comparable magnitudes today. 

It is conceivable that $\rho_v$ is associated with 
a false vacuum energy, with the true vacuum possessing a zero cosmological 
constant \cite{4, 5, hall}. In this
admittedly modest approach to the problem, one tries to identify the origin
of $\rho_v$ and also ensure that the false vacuum is sufficiently long lived.
To this we wish to add in this paper an important new ingredient, namely
inflation. This would help us explain how the universe got stuck in the
false vacuum in the first place.

The model described below is organized within the framework of
supersymmetric
hybrid inflation \cite{hybrid} which is associated with the breaking of 
some gauge symmetry
$G$ to $H_0$, where $H_0$ could be the MSSM gauge group or something larger. A
remarkable feature of these models is that the symmetry breaking scale  
of $G$
is estimated from the quadrupole anisotropy, 
$\frac{\delta T}{T}$, to be of order $10^{16}$ GeV, 
the supersymmetric GUT scale, $M_{GUT}$. A nice,
and perhaps the simplest, example of $G$ is the MSSM gauge symmetry
supplemented by a gauged $U(1)_{B-L}$ symmetry \cite{u1}. 
To realize $(\rho_v )^{1/4} \sim 10^{-3}$ eV we assume, following 
\cite{hall}, that the 
fundamental supersymmetry breaking scale in nature is $\sim$ TeV \cite{pomerol}, 
so that the 
gravitino mass $m_{3/2} \sim$ TeV$^2$/$M_P$ more or less coincides with
$(\rho_v)^{1/4}$. Furthermore, following \cite{hall}, a new (acceleressence) 
sector containing a chiral
superfield $\chi$ is introduced, which communicates with other 
sectors only via gravity. The $\chi$ sector will be arranged to yield a
potential
which has a false (metastable) minimum separated by $\rho_v$ from the true
minimum with zero cosmological constant.

We will see that during inflation driven from the visible sector, taking
supergravity corrections into account, 
the scalar component of $\chi$ acquires 
a mass of order the Hubble constant $H$, causing it to be trapped in the false 
minimum at the origin. If 
the barrier separating the two minima is sufficiently high, the field stays
stuck in the false vacuum even after inflation ends.
Because the gravitino is ultralight, the MSSM sector does not provide a
suitable
cold dark matter (CDM) candidate. 
Potential CDM candidates include stable relics from the supersymmetry 
breaking sector \cite{hall}, or a suitable pseudogoldstone boson \cite{nasri}, 
and finally axions that we shall shortly discuss.

The model consists of three components namely, the visible sector, 
a strongly coupled supersymmetry breaking hidden sector, and the acceleressence sector 
which we will refer to as $G, T$ and
$\chi$ sectors respectively. The $G$ sector, as we shall see, consists of the 
MSSM superfields and additional ones used to implement inflation and 
the axion mechanism.
We do not need to specify the details of the supersymmetry 
breaking sector except 
to note that it contains a (possibly composite) chiral field $T$, whose
auxiliary component has a vev $\langle F_T \rangle \sim $ TeV$^2$. 
The $T$ sector communicates 
via gauge interactions with the visible sector, so that the supersymmetric 
partners 
of the known (SM) particles can acquire masses in the range of $M_Z$ to TeV. 
The $\chi$ sector, following \cite{hall}, allows us to relate the observed 
vacuum energy density to a false vacuum energy density. As stated before,  
this sector consists of a chiral superfield $\chi$ which communicates with 
the two sectors $G$ and $T$ only via gravity. With the superpotential
\be
W_{acc} = \frac{\sigma}{3}\chi^3,
\ee
and including soft supersymmetry breaking terms, the $\chi$ potential takes 
the form 
\be
V_{acc} = \sigma^2 |\chi|^4 - (A \chi^3 + h.c.) + m^2 |\chi|^2 + V_{1},
\label{vchi}
\ee
where $\sigma, A$ can be made real and positive by proper phase rotations 
of the fields.
Here, both $A$ and $m$ are of order $10^{-3}$ eV, 
and $V_{1}$ is adjusted
to make the total energy density vanish at the absolute minimum which lies
at $\chi = \frac{3A + \sqrt{9A^2 - 8 \sigma^2 m^2}}
{4\sigma^2}$ for $9A^2 > 8 \sigma^2 m^2$. Note that $V_{acc}$ also has a local (false)
minimum at $\chi = 0$ 
which is separated from the true minimum by $\rho_v$.
It is possible to make the lifetime of this
metastable state (much) greater than the age
of the universe. 
The dark energy conundrum could be explained  
if the field $\chi$ is trapped at the origin rather than 
in the true minimum. We will show that supersymmetric hybrid inflation 
provides a natural mechanism to drive the $\chi$ field to the false minimum 
thereby realizing the acceleressence scenario. 

The $G$ sector contains the superpotential responsible for 
the simplest model of hybrid inflation \cite{hybrid, nefer}
\be
W_{inf} = \k S [\phi \bar \phi - M^2],
\label{1}
\ee
where $\phi, \bar \phi$ denote a conjugate pair of non-$G$ singlet 
superfields, $S$ is a gauge 
singlet superfield and a $U(1)_R$ symmetry is imposed under which 
$S \rightarrow e^{i\alpha} S$, $\phi \bar \phi 
\rightarrow \phi \bar \phi$, and $W_{inf} \rightarrow  
e^{i\alpha} W_{inf}$. 
The parameters $\k$ and $M$ can be made real and positive by 
field redefinitions. In the unbroken supersymmetric limit, 
vanishing of the $F$- and $D$-terms imply that the 
supersymmetric vacuum corresponds to $\langle S \rangle = 0, {|\langle \bar 
\phi \rangle}| = |\langle  
\phi \rangle | \equiv M$.
To realize inflation, $S$ is displaced from its present day location to values that exceed $M$. The appearance of a vacuum energy density of order 
$ \k^2 M^4 $ induces radiative corrections to the tree level potential, with 
the result that $\frac{\delta T}{T} \propto (\frac{M}{M_P})^2$ \cite{hybrid, 
nefer}. Thus, $M$ is of order $10^{16}$ GeV, the supersymmetric GUT scale \cite{hybrid}. The 
scalar spectral index in this class of models is estimated to be 
$n_s = 0.99 \pm 0.01$ \cite{hybrid, nefer}. 

Let us now include supergravity corrections that link the inflaton and 
the $\chi$ sector. The supergravity corrections 
coming from supersymmetry breaking in the strongly-coupled sector 
are small during inflation and would only play a significant role near 
the end of inflation, by which time the $\chi$ field is trapped in  
the false minimum.
Assuming minimal supergravity, the scalar potential corresponding to a 
superpotential $W$ and K\"ahler potential $K$ is given by \cite{WB} 
\bea V = {\mathrm{exp}}\left(\frac{K}{M_P
^2}\right)\left[ \left(W_i + \frac{K_i W}{M_P ^2}\right)K_{ij^*}
^{-1}\left(W^*_{j^*} + \frac{K_{j^*} W^*}{M_P ^2}\right) -3
\frac{|W|^2}{M_P ^2}\right],
\label{Vsc}
\eea 
where $K_i = \pa{i}K$,
$W_i = \pa{i}W$, $K_{ij^*} ^{-1}$ is the inverse of the K\"ahler
metric and the indices $i,j$ run through all chiral fields.

We can parametrize,
without explicit details of the supersymmetry breaking sector, the
supergravity mediated supersymmetry breaking effects on the
visible and $\chi$ sector by explicitly including a constant term
$W_0$ in the superpotential. The presence of $W_0$ 
ensures the cancellation of the cosmological constant so that the vacuum 
energy at the global minimum is zero. 
The size of supersymmetry breaking in the $T$ sector implies that 
$W_0 \simeq m_{3/2} M^2_P \sim O ({\mathrm{TeV^2}}) M_P$ and 
$\langle W_i K^{-1}_{ij^*} W^*_{j^*} \rangle \sim O (\mathrm{TeV}^4)$ 
to leading order in $1/M_P$ (provided there are no Planckian vevs).

With the minimal K\"ahler potential $K_1 = S S^{\dag} + \phi \phi^{\dag} 
+ \bar \phi \bar \phi^{\dag}$ from the inflationary sector and 
$K_{2} = \chi \chi^{\dag}$ from the acceleressence sector,
the scalar potential is given by (we employ the same notation for superfields
and their corresponding scalar components)  
\bea
V & = & {\mathrm{exp}}\left(\frac{K_1 + K_2}{M^2_P}\right) 
\biggl [ |\k S \bar \phi + \phi^{*} \frac{W}{M^2_P}|^2 + |\k S \phi  
+ \bar \phi^{*} \frac{W}{M^2_P}|^2 + |\k (\phi \bar \phi - M^2) 
\nonumber \\
& & + S^* \frac{W}{M^2_P}|^2
+ |\sigma \chi^2 + \chi^* \frac{W}{M^2_P}|^2 + ... -3 \frac{|W|^2}{M^2_P}
\biggr ],
\label{v}
\eea  
where $W = W_{inf} + W_{acc} + W_{MSSM} + W_0$, and the 
ellipsis represent contributions from the MSSM fields. With 
$|\bar \phi| = |\phi|$ along the $D$-flat direction
of the scalar potential, the
tadpole term $-2\k M^2 m_{3/2} S + h.c.$ induces a shift in the
vevs \cite{svev}:
\be
\langle S \rangle \simeq \frac{m_{3/2}}{\k}; ~~ |\langle \phi \rangle| =
|\langle \bar \phi \rangle| \simeq M(1 - \frac{m^2_{3/2}}{2 \k^2 M^2}). 
\ee
The corresponding $F$-terms are
\be
F_S \simeq -\frac{m^2_{3/2}}{\k}; ~~ F_{\phi} = F_{\bar \phi} \simeq m_{3/2}M.
\label{fs}
\ee

The supergravity corrections play an important role during inflation. With 
$\phi = \bar \phi = 0$ and $|S| > M$, the scalar potential  
is given by  
\be
V \simeq \k^2 M^4 \left [1 + |\frac{\chi}{M_P}|^2 \right ] - 
\left ( \frac{\sigma \k M^2}{3 M_P} 
\frac{S^*}{M_P} \chi^3 + h.c. \right ) + \sigma^2 |\chi|^4 ,
\ee
where only the dominant lower order terms are displayed, and the higher order terms in $\chi$ can be safely ignored for our discussion. 
Note that during inflation, the $\chi$ field acquires a positive mass 
squared larger than
$H^2$ ($\sim \frac{\k^2 M^4}{3 M^2_P}$). The coefficient of 
$\chi^3$ term, $\frac{\sigma}{\sqrt{3}} \frac{S}{M_P} H$, is suppressed 
compared to $H$, and therefore 
$\chi$ rapidly settles at the origin during inflation.

With the end of inflation, the effective potential for $\chi$ is given by 
Eq.(\ref{vchi}) which can be seen as follows. 
The soft mass squared term $m^2_0 |\chi|^2 = a m^2_{3/2} |\chi|^2$, where $a \sim O(1)$, 
arises from $W_0$ introduced to cancel the cosmological constant as discussed 
earlier, with $m^2_{3/2} \sim O$(meV$^2$). 
Terms of $O(m_{3/2}) \chi^3$ do not follow in the same way 
because of a cancellation between contributions from 
$W_{\chi} K^{-1}_{\chi \chi^*} K_{\chi^*} \frac{W^*}{M_P ^2}$ 
and $-3 \frac{|W|^2}{M_P ^2}$ terms.  
With the minimal K\"ahler potential,
 given that the inflationary sector contains the 
 vevs $ |\langle \phi \rangle | = \langle \bar{\phi} \rangle | \simeq M_{GUT}$, we find the term 
$O(m_{3/2} (\frac{M_{GUT}}{M_P})^2) \chi^3 + h.c.$.
To realize a $\chi^3$ term of the correct magnitude, we include 
the higher order K\"ahler term \cite{hall} 
\be
\int d^4 \theta ~\frac{T + T^{\dag}}{M_P} \chi^{\dag} \chi ,
\ee
from which the term $A \chi^3 $ in Eq.(\ref{vchi}) can be 
generated, where $A \sim 
\sigma \frac{F_T}{M_P} \sim \sigma 10^{-3}$ eV. 
As for the quartic 
term, it just comes from the usual $F$-term squared, i.e. $W_i
K_{ij^*} ^{-1} W^*_{j^*}$.
Thus after inflation, the $\chi$ sector scalar potential takes the form  
\bea
V_{acc} = \sigma^2 |\chi|^4 - \left [(A + O(m_{3/2})
\left(\frac{M_{GUT}}{M_P} \right)^2 )\chi^3 + h.c. \right ]+ m^2_{0} |\chi|^2 
+ V_{1},
\eea
which is essentially equivalent to Eq.(\ref{vchi}). 

The next question we would like to address is that of dark
matter. 
The superlight gravitino with mass $\sim 10^{-3}$ eV is not a 
suitable dark matter candidate which forces us to look for 
alternative CDM candidates. 
One plausible candidate would be the lightest field in the
supersymmetry breaking hidden sector as one would expect it to
have quantum numbers not shared by fields in the other sectors and
hence, be stable \cite{hall}.
Another plausible candidate could be a pseudogoldstone boson such as 
the majoron, associated with a spontaneously broken global $U(1)_{B-L}$ 
symmetry \cite{nasri}. 
We will focus here on axion CDM introducing a $PQ$ symmetry 
$U(1)_{PQ}$ \cite{pq}, 
since the associated physics can also be exploited to 
resolve the MSSM $\mu$ problem \cite{khalil,ls}. 
Implementation of this mechanism turns out to be not entirely straightforward.

The axion mechanism is easily implemented in models 
in which the gravitino mass, $m_{3/2} \sim $ TeV. 
With the introduction of two $G$-singlet superfields $N, \bar N$ 
carrying appropriate ${PQ}$ and $R$ charges, the superpotential 
terms $N^2 \bar N^2 / {M_P}$ and $N^2 H_u H_d / {M_P}$  
can provide ($H_{u}, H_d$ denote the MSSM higgs superfields) a
vev for the scalar components of $N, \bar N$ of magnitude 
$(m_{3/2} M_P)^{1/2}$, after taking the supersymmetry 
breaking terms (proportional to $m_{3/2}$) into account. 
This vev has the right order of magnitude ($\sim 10^{11}$ GeV) 
for axion dark matter,  
assuming that $m_{3/2} \sim$ TeV $\sim m_N$ ($m_N$ is the soft mass for $N$). 
The second field 
$\bar N$ is needed to ensure the invariance of the superpotential, under 
$U(1)_{PQ}$. Its vev breaks $U(1)_R$ and ensures that the $R$-axion is 
phenomenologically harmless.  

With $m_{3/2} \sim 10^{-3}$ eV in our present case, the above scenario 
cannot be realized in the simple way outlined above. 
Furthermore, superpotential terms such as $N^2 \bar N^2 /{M_P}$ give rise 
to $F$-term contributions $\gg$ TeV$^2$, which can be disastrous for 
the $\chi$ sector, through non-minimal K\"ahler terms such as 
$\int d^4 \theta N^{\dag} N \chi^{\dag} \chi /M^2_P$. 
We will attempt to implement the axion 
mechanism with a single $G$-singlet superfield $N$, by retaining only the 
superpotential term 
\be 
W_{PQ} = \lambda \frac{N^2 H_u H_d}{M_P},
\label{wn}
\ee
and letting $m_N$, the coefficient of the mass term associated 
with the real component of $N$, also called the saxion,   
be a free parameter to be determined from the consistency 
requirements. Namely, that the $\mu$ problem is resolved with a $N$ vev  
of order $10^{11}$ GeV in order to generate axion dark matter \cite{khalil}, 
and  
that there are no cosmological problems 
associated with the $N$ field. How $m_N$ acquires the desired mass scale 
requires a more complete analysis of supersymmetry breaking which is beyond the scope 
of this paper. The cosmological evolution of the saxion field turns out to be 
somewhat non-trivial.  
The $R$ and $PQ$ charges of the various superfields are 
listed in Table \ref{tab1}.

\begin{table}[h]
\begin{center}
\begin{tabular}{c | ccccccccccc}
   Field& $S$ & $\phi$ & $\bar \phi$ & $H_{u,d}$ & $Q$ & $U^c$ & $D^c$ & $L$ & $E^c$ & $N$ & $\chi$ \\
  \hline
 $R$& 1& 0 & 0 & 1/2 & 1/2 & 0 & 0 & 1/2 & 0 & 0 & 1/3 \\
 $PQ$& 0 & 0 & 0& 1 & -1 &0 &0 &-1 &0 &-1 & 0 \\
\end{tabular}
\end{center}
%\caption{\label{TabCharge} {\small $R$ and $PQ$ charge assignments for
\caption{\small $R$ and axion ($PQ$) charge assignments for
various superfields. We have used the convention under which $[W]_R =1$. Additionally, the
fields $Q$, $L$, $E^c$, $U^c$ and $D^c$ are odd under a $Z_2$ matter parity to
eliminate rapid proton decay.}
\label{tab1}
\end{table}

The potential responsible for breaking the axion symmetry is taken to be
\be
V_{PQ} \simeq -m^2_{N} |N|^2 + \lambda^2 (\frac{M_{W}}{M_P})^2  |N|^4 + V_{2},
\nonumber 
\label{vn}
\ee
where a negative mass squared term for the $N$ field may, for instance, be
induced via radiative corrections \cite{ls}. The second term follows from the 
superpotential in Eq.(\ref{wn}) after electroweak symmetry breaking.  
A constant term $V_{2}$ has been included to set  $V_{PQ}$ to zero 
at the true minimum.
Requiring
$ f_a = |\langle N \rangle| \sim 10^{11}$ GeV \cite{axion}  
yields $m_N \sim \lambda 
\times  10^{-5}$ GeV $\sim 10^{-7}$ GeV, with $\lambda \sim 10^{-2}$ so that 
the $\mu$ term $\sim$ 100 GeV. 
The saxion mass then is also of this magnitude. 
Since we have a very light and consequently a long lived (essentially stable) 
scalar we should ensure that no cosmological difficulties arise
as a consequence. 
Note that in Eq.(\ref{vn}) we could introduce an additional quartic term 
$\gamma |N|^4$, with $\gamma \sim 10^{-38}$. This latter coupling, whose origin 
like that of $m_N$ we will not discuss here, will be useful in 
cosmology. The values for $m_N$ and $\gamma$ proposed here suggest the presence of heavy fields that link the $N$ superfields with the supersymmetry breaking 
sector. 

In contrast to the $\chi$ field that remains trapped at the origin both
during and after inflation, 
the saxion field must reach its minimum to implement proper 
breaking of the $U(1)$ axion symmetry. 
In principle, it could stay at the origin during inflation. However, 
axion models are often plagued by the domain wall problem \cite{domain} 
and we prefer to
circumvent this by letting $N$ roll away from the origin during inflation. This
can be accomplished by introducing suitable non-minimal K\"ahler potential 
terms.
Consider, for instance, the K\"ahler potential
\be
K_1 + \k_1 \frac{N N^{\dag} S S^{\dag}}{M^2_P}, 
\ee
so that, during inflation, the relevant potential involving the 
$N$ field is given by
\bea
V_{PQ,~{inf}} &\simeq & -(3\beta H^2 + m^2_N)|N|^2 + 
[ \frac{3}{2}(1 + 2\beta + 2\beta^2)\frac{H^2}{M^2_P} + \gamma ] |N|^4 
\nonumber \\
& & + 3 \beta^2 \frac{H^2}{M^2_P} |N|^2 |S|^2 + ... ,                             
\eea
where $\beta = (\kappa_1 - 1) > 0$. For $\beta \lesssim 10^{-1}$ , the field $N$ 
is rapidly driven to $ \sqrt{\beta}M_P$. Note that the induced mass-squared 
term for $S$ is suppressed
relative to $H^2$ by a factor of $\beta^3$, so that the inflationary scenario
described earlier remains intact.
As the Hubble induced mass drops below $m_N$ after reheat, which happens at a
temperature of order $10^5$ GeV, the $N$ field moves, because of the quartic 
term $\gamma |N|^4$, to a new minimum at around
$10^{13}$ GeV. A further drop in temperature to  
$10^2$ GeV leads to the appearance of electroweak vevs, in which case 
the potential in Eq.(12) effectively takes
over, and the $N$ field reaches its true minimum value of around $10^{11}$ GeV.
This creates a cosmological problem since the energy stored in the $N$ field 
($\sim \lambda^2 \times 10^{12}$ GeV$^4$) is comparable 
to the radiation energy density ($\sim 10^8$ GeV$^4$) and,
with $N$ having a lifetime that far exceeds the age of the universe, 
$N$ would become the dominant component in the universe.

One mechanism for overcoming this is to invoke an epoch of thermal
inflation \cite{thermal}. 
We will not provide any details here since a similar problem was
encountered in \cite{hall2} where the decay of a heavy particle 
was employed to dilute sufficiently the saxion energy density. 
Of course, the release of entropy also
dilutes any pre-existing baryon asymmetry and a mechanism should be found to
resolve this problem \cite{shahida}.
Finally, let us note that in the presence of axions, the gravitino is replaced
by the axino, with mass $\sim 10^{-7}$ eV (for $\lambda \sim 10^{-2}$), 
as the LSP. Its contribution  
to the energy density of the universe, like the gravitino, is negligible. 
Cold dark matter comes from axions and possibly also the saxion.

Some remarks about the $R$-axion are in order here. The $U(1)_R$ symmetry 
is explicitly broken by the constant superpotential term $W_0$. With a 
superpotential $W_0 + W_1$, where $W_1 = W_{acc} + W_{inf} + 
W_{PQ} + W_{MSSM} + W_{hidden}$, the $R$-axion mass is estimated 
to be \cite{bagger}
\be
m^2_a = \frac{8}{f^2_R} \frac{W_0 |\langle W_{1i} K^{-1}_{ij^*} K_{j^*} - 
3 W_1 \rangle |}{M^2_P} ,
\label{a}
\ee 
where the $R$-axion decay constant $f_R \sim r_i r_j v_i v^*_j \langle K_{ij^*} \rangle 
$, and $r_i$ and $v_i$ are the $R$ charges and vevs 
of the fields respectively. With the large $R$-singlet vev of $|\langle \phi 
\rangle | =| \langle \bar \phi \rangle| \simeq   
M$ and hidden sector fields (generically labeled $T$) with vevs $\langle T 
\rangle \sim \sqrt{\langle F_T \rangle} \lesssim O$(TeV), we expect that 
\be
f_R \sim O({\rm TeV}),
\label{a1}
\ee 
\be
|\langle W_{1i} K^{-1}_{ij^*} K_{j^*} - 3 W_1 \rangle | \sim 
\langle W_{1i} \phi \rangle 
\sim m_{3/2}M^2.
\label{a2}
\ee
Substituting Eqs.(\ref{a1}) and (\ref{a2}) in Eq.(\ref{a}), we obtain an 
$R$-axion 
mass of $\sim$ 10 GeV which is consistent with the astrophysical 
constraints.

In conclusion, we have explored a scenario in which supersymmetric 
hybrid inflation could 
play an essential role in understanding the origin of dark energy. 
Even though the true vacuum has a zero cosmological constant (how this comes about is beyond the scope of this paper), supergravity corrections
during inflation can trap acceleressence field at the origin, 
which happens to be a
local (false) minimum. The energy density scale separating the true
vacuum from the false one is arranged to be of order TeV$^2$/$M_P \sim 
10^{-3}$ eV. 
Because of the low ($\sim$ TeV) fundamental supersymmetry breaking scale, 
the MSSM 
LSP is not a plausible cold dark matter candidate. 
There are three potential CDM candidates including axions. 
It turns out that in addition to the axions, 
the saxion may also be a significant component of cold dark matter. \\

We thank V.N. \c{S}eno\u{g}uz for fruitful discussions. 
Q.S. would like to thank 
W. Buchmuller for useful correspondence regarding the superlight gravitino 
and R.N. Mohapatra for pointing out that the Majoron in \cite{nasri} is a 
possible cold dark matter candidate. A.S. thanks E. Poppitz for 
valuable correspondence regarding $R$-axion. S.P.N. thanks Hock-Seng Goh 
for helpful discussions.
This work is supported by DOE under contract number DE-FG02-91ER40626.

\end{document}